# Bulk carrier lifetime surpassing 600 μs in Upgraded Metallurgical-grade Silicon multicrystalline wafers after Phosphorus Diffusion Gettering

*Nerea Dasilva-Villanueva\*, Sergio Catalán-Gómez, David Fuertes Marrón, Miguel García-Corpas, Carlos del Cañizo*

N. Dasilva-Villanueva, S. Catalán-Gómez, D. Fuertes Marrón, M. García-Corpas, C. del Cañizo
Instituto de Energía Solar, Universidad Politécnica de Madrid, ETSI Telecomunicación, Av. Complutense 30, E-28040 Madrid (Spain)
Email: nerea.dasilva@ies.upm.es

Upgraded metallurgical-grade (UMG) Si is obtained via a purification route alternative to the one used for conventional polysilicon and with significantly reduced environmental impact. Additionally, despite a lower purity level in the feedstock than polysilicon, UMG-Si has demonstrated potential for the fabrication of highly efficient and low-cost solar cells. Low initial bulk carrier lifetimes recorded in UMG-Si bare wafers can be improved by means of an adequate Phosphorus Diffusion Gettering (PDG) process to the level of mc-Si. In this letter, optimized PDG processes for UMG-Si are reported, resulting in increased values between 20 and 250 times the original carrier lifetimes and record figures above 645 μs.

## 1. Introduction

Throughout the last decade, researchers from the photovoltaic (PV) field have focused efforts on the upgrading of silicon material for solar cells. In a scenario in which the aim is not only the reduction of the levelized cost of electricity, but also the minimization of $CO_2$ emissions, upgraded metallurgical-grade (UMG) Si, a material obtained via an alternative purification route to that of conventional polysilicon, is called to play a key role since it results in lower manufacturing cost and environmental impact in comparison to polysilicon[1–3], demonstrating at the same time similar efficiencies (with values up to 20.76%) in Al-BSF and PERC architectures.[4–7]

Notwithstanding, it is well known that UMG-Si feedstock has a high impurity concentration (metallic[8] and non-metallic such as oxygen precipitates[9]), and typically shows certain carrier-trapping effect [10] and a moderate doping compensation character.[11] All these aspects can affect the dominant carrier recombination mechanisms. In fact, bulk minority carrier lifetime values of this material are typically low, in the range of 5-10 μs[4,12] for untreated wafers. Such low initial lifetime values could limit the cell efficiencies, as predicted by simulations[13,14], and confirmed in the case of monocrystalline, Czochralski (Cz)-grown solar cells.[5] Consequently, the implementation of a Phosphorous diffusion gettering (PDG) process is mandatory in order to improve the minority carrier lifetime within the material.[15]

Regarding the type of crystallinity, very high lifetime values after PDG processes have been reported for monocrystalline Czochralski n-type UMG-Si[5,6,11,16,17], reaching values in the range of miliseconds. However, when analyzing multicrystalline (mc) UMG-Si material, only modest lifetime values have been achieved[9,10,12,16,18–20], with most studies reporting values around 70 μs.[10,21] On the other hand, comparing literature data of mc-Si purified by different ways, it can be observed than the efficiency of PDG for boosting the carrier lifetime has been clearly higher for conventional electronic-grade Si than for UMG-Si.[22] This suggests that the PDG optimal conditions for mc-UMG material may have not been established yet or alternatively, that the specificity of each type of UMG-Si requires its own optimized PDG process.

In this work we report what is to the best of our knowledge, the highest carrier lifetime measured so far in mc-Si UMG wafers, obtained after a PDG process which is compatible with high-efficiency solar cell processing. We also investigate the convenience of performing a consecutive second PDG treatment and ascertain its optimal conditions.[23,24] As it will be shown, the success of such gettering processes is related with the initial state of the bare wafers.

## 2. Results and Discussion

Effective lifetime curves for a representative batch are represented in Figure 1, showing passivated measurements for samples before the PDG process, as well as after the two P-diffusions carried out. A substantial improvement is shown after the first gettering step, with values ranging between 200 up to 500 microseconds at an injection level of $10^{15}$ cm$^{-3}$, a raise that accounts for increases up to two orders of magnitude in some samples.

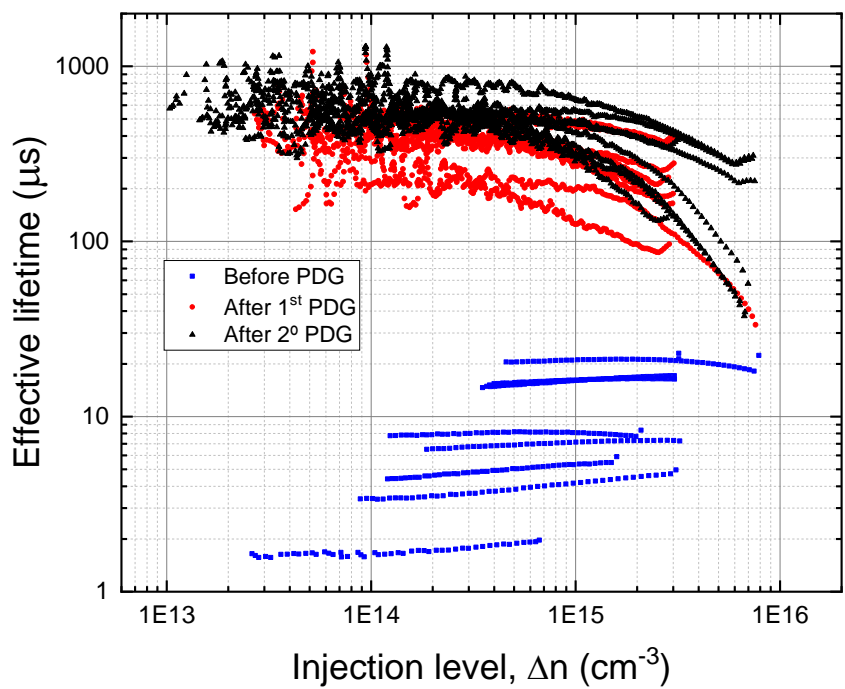

**Figure 1.** Effective lifetime curves of eight passivated samples analyzed as a function of the injection level, before PDG (blue squares), after a PDG process (red circles) and after two PDG processes (black triangles). Passivation was carried out with a 0.1M IE solution.

In the literature, the presence of thermal donors as oxygen interstitials as well as oxygen precipitates has been related to inefficient or even detrimental PDGs[9], in terms of carrier lifetime in UMG material.[25] For this reason, FTIR studies were carried out, both before and after PDG, in order to identify O-related fingerprints that could affect the lifetime. FTIR measurements for an exemplary batch are shown in Figure 2. No changes were observed in FTIR spectra after the second PDG in any of the samples. Transmittance is plotted as a function of the wavenumber before (blue squares) and after (red circles) the first PDG process. The spectra are represented in the range from 900 to 1300 cm-1 where the oxygen-related bands are expected.[26] A first band would be expected at 1107 cm$^{-1}$, ascribed to the antisymmetric stretching of interstitial oxygen atoms in the quasi-molecule Si-O$_i$-Si, and a second band at around 1230 cm$^{-1}$, corresponding to platelet precipitates.[27] As it can be seen, no traces of oxygen, neither in precipitate nor interstitial form, are found in the samples. Notwithstanding, in our previous work traces of oxygen precipitates were found before PDG processing in UMG samples that showed just a modest lifetime improvement upon the first gettering step (in the range of 100 μs).[28]

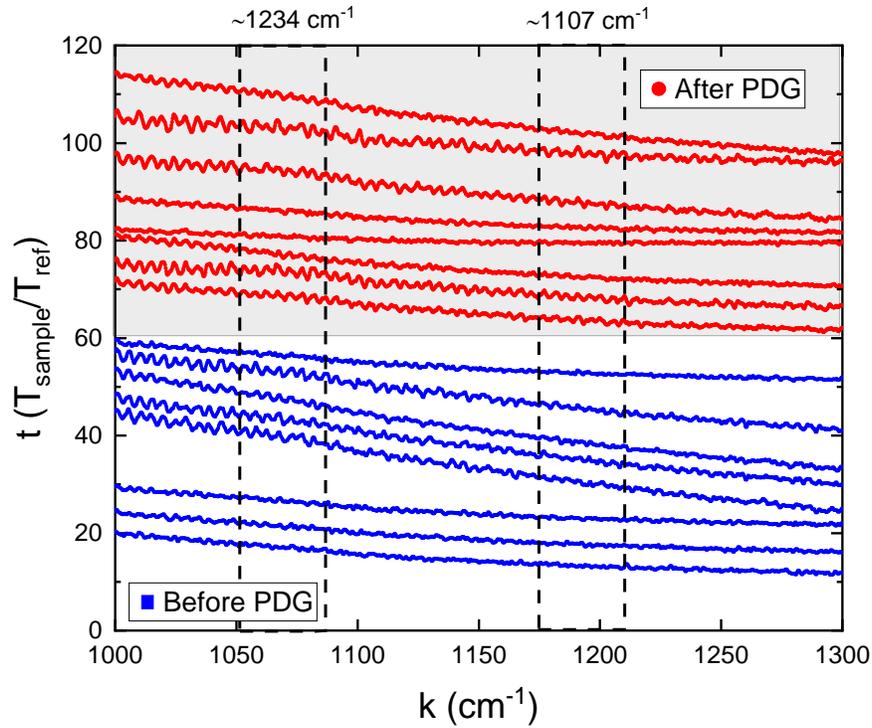

**Figure 2.** Relative transmittance (sample over reference transmittance) FTIR measurements in the range 1000-1300 cm$^{-1}$ for samples before (blue) and after (red) PDG.

The possibility of further improvement, as well as the eventual reduction of the somewhat wide dispersion of improvement ratios recorded between samples, raise the question of the potential benefit of performing a second PDG process on the same samples. As seen in Figure 1, after the second gettering step, although more modest, further improvements are recorded, ranging from 1 to 2.5 times the post first PDG lifetime and resulting in overall improvements between 20 and 250 times the original effective lifetime, raising lifetimes up to a maximum recorded value of 645 µs. In order to increase the statistical significance of the study, both processes were carried out on 64 samples. In Figure 3, the effective lifetime distributions after first and second PDG are shown for all wafers at an injection level of $10^{15}$ cm$^{-3}$. Each point represents a single wafer. Dashed connection lines are also included in the graph to keep track of the change of lifetime values after the second PDG. It can be observed that the second PDG has an overall benign effect in most of the cases, proving its usefulness.

As shown in Figure 3, different conditions have been analyzed for the second gettering process, varying both its time and temperature. The optimal conditions used for the first PDG, 780ºC during 60 minutes, were optimized and established previously [10,28]; and are different from those needed to optimize a P-emitter. The fact that the second PDG process can be carried out at higher temperatures and shorter processing times is beneficial, as the diffusion of the P-emitter could further improve the result of the first gettering.

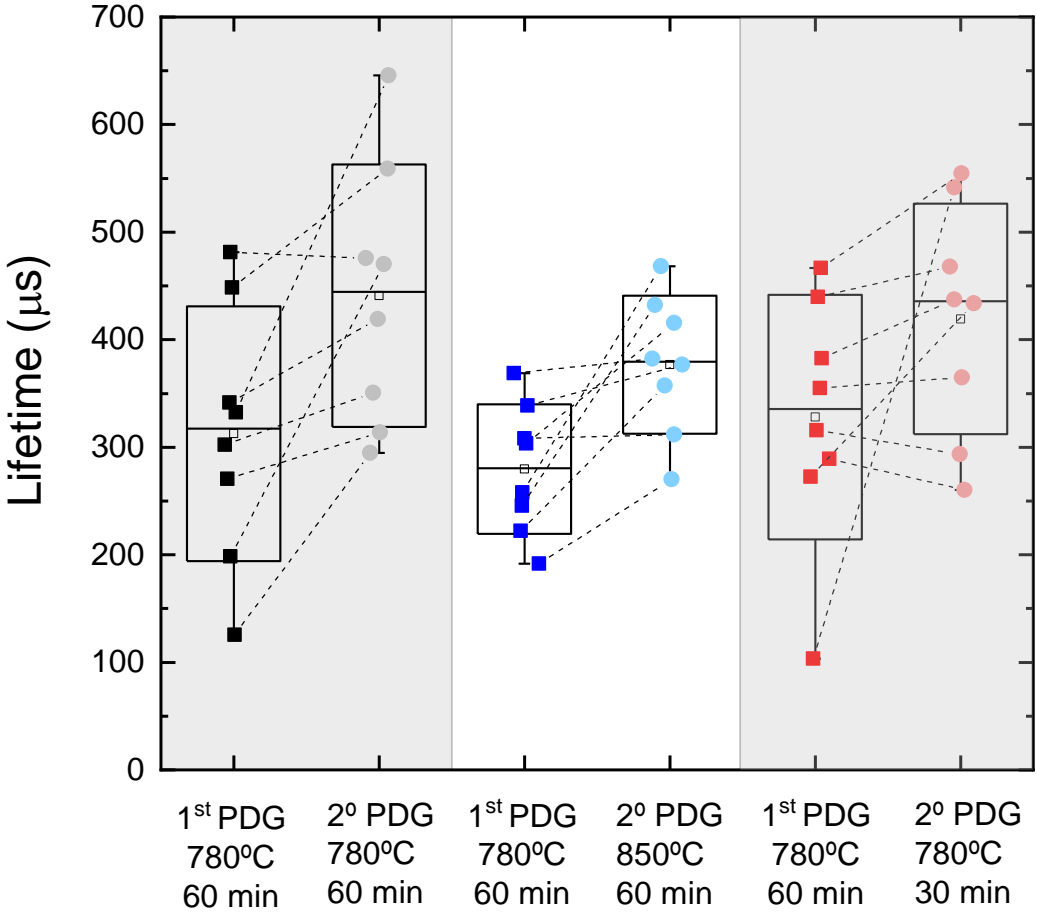

**Figure 3.** Effective lifetime variation after one and two PDG processes for different 2nd PDG conditions. In the box-type plot whiskers represent max-min variation, the height of the box represents standard deviation, the middle line represents the median and the average value is indicated with an open square. Connection dashed lines are included to identify the change of individual samples.

## 3. Conclusions

Effective lifetime values up to 645 µs in multicrystalline UMG-Silicon samples are obtained via optimization of two consecutive Phosphorus Diffusion Gettering processes. Total improvements of effective carrier lifetimes after PDG range between 20 and 250 times the starting values of bare wafers, resulting in the highest measured values up-to-date for this type of material.

No presence of either interstitial oxygen or oxygen precipitates is observed in the samples analyzed, which could be related with the lifetime improvement observed after the diffusion process.

The positive response to the material to PDG is thus proved, and the usefulness of a second process is also demonstrated, as further improvements are observed under a relatively wide range of processing conditions. These results confirm the potential of UMG-Si as a highly efficient and environmentally friendly alternative feedstock to conventional poly-Si for the fabrication of Si-based PV.

## 4. Experimental section

UMG material has been manufactured via the metallurgical route by FerroGlobe, and crystallized as multicrystalline ingots.[4] The resulting ingots were sliced in 15x15 cm$^2$ wafers of 200±5 µm thickness with resistivity values ranging from 1.2 to 1.5 Ω·cm. The compensation level of the UMG material used in this study and defined as $(N_A+N_D)/(N_A-N_D)$, where $N_A$ and $N_D$ are the concentrations of p-type and n-type dopants, respectively, varies along the ingot between 2 and 4. Each wafer is cut in 5x5 cm$^2$ samples for further processing.

Minority carrier lifetime data were obtained from bare wafers by quasi-steady state photoconductance (QSSPC) measurements and transient measurements with the Sinton Instruments WCT-120 tool, using a slow flash decay mode with a constant decay of 3.6 ms and a fast flash decay mode with a constant decay of 23 µs, respectively; as measured at the reference cell. For the acquisition of the lifetime curves, an optical constant of 0.7 and the Dannhäuser carrier mobility model were used. While the model more adequate for measurements in compensated materials is the Klaassen one[29], for the level of compensation present in our material and the level of injection analyzed, the differences are not particularly relevant and Dannhäuser's is kept to avoid calculation errors[30]. This characterization was performed before and after 0.1M iodine-ethanol (IE) passivation for the reduction of surface recombination. After a chemical etching with CP4 ($HNO_3$/HF) and RCA1 surface cleaning, the wafers were subjected to a thermal treatment. PDG processes were performed in a tubular furnace for different times in $O_2$ and $N_2$ atmosphere. Liquid $POCl_3$ was the P source during the process at a fixed temperature of 780ºC. After PDG the emitter formed atop the surface was removed by CP4 etching. Then, PCD measurements were again taken without and with IE passivation and converted to lifetime curves.

To test the viability of further lifetime improvements, a second PDG and subsequent etching and characterization were carried out following the same steps.

All samples, both before and after PDG, were studied by means of Fourier transform infrared (FTIR) spectroscopy (Thermo Scientific Nicolet iS10) in the 400-4000 $cm^{-1}$ range in transmission mode, with the aim of detecting the fingerprint of O-related impurities in the Si crystal. A float zone p-type Si sample of 200 μm thickness and 3-5 Ω·cm resistivity was used as a reference sample.

## Acknowledgements

This work is part of the R&D SOLAR-ERA.NET Cofund project CHEER-UP (PCI2019-111834-2), funded by MCIN/ AEI/10.13039/501100011033/ and the European Union, and also part of the Project MADRID-PV2 (S2018/EMT-4308) funded by the Regional Government of Madrid with the support from FEDER Funds. S.C.G acknowledges grant FJC2019-041616-I funded by MCIN/AEI/ 10.13039/501100011033. Juan José Torres, Bo-Kyung Hong and Manuel Funes are acknowledged for support in wafer processing and for fruitful discussions. Aurinka PV is acknowledged for UMG wafer supply.

S. Catalán-Gómez currently at Institute for Optoelectronics Systems and Microtechnology, ISOM, Universidad Politécnica de Madrid, ETSI Telecomunicación, Avda. Complutense 30, E-28040 Madrid (Spain).